# Copied citations create renowned papers?

M.V. Simkin and V.P. Roychowdhury
*Department of Electrical Engineering, University of California, Los Angeles, CA 90095-1594*

**Abstract.** Recently we discovered (cond-mat/0212043) that the majority of scientific citations are copied from the lists of references used in other papers. Here we show that a model, in which a scientist picks three random papers, cites them, and also copies a quarter of their references accounts *quantitatively* for empirically observed citation distribution. Simple mathematical probability, not genius, can explain why some papers are cited a lot more than the other.

During the "Manhattan project" (the making of nuclear bomb), Fermi asked Gen. Groves, the head of the project, what is the definition of a "great" general [1]. Groves replied that any general who had won five battles in a row might safely be called great. Fermi then asked how many generals are great. Groves said about three out of every hundred. Fermi conjectured that considering that opposing forces for most battles are roughly equal in strength, the chance of winning one battle is ½ and the chance of winning five battles in a row is $1/2^5 = 1/32$. "So you are right General, about three out of every hundred. Mathematical probability, not genius." The existence of military genius was also questioned on basic philosophical grounds by Tolstoy [2].

A commonly accepted measure of "greatness" for scientists is the number of citations to their papers [3]. For example, SPIRES, the High-Energy Physics literature database, divides papers into six categories according to the number of citations they receive. The top category, "Renowned papers" are those with 500 or more citations. Let us have a look at the citations to roughly 24 thousands papers, published in Physical Review D in 1975-1994 [4]. As of 1997 there where about 350 thousands of such citations: fifteen per published paper on the average. However, forty-four papers were cited five hundred times or more. Could this happen if **all papers are created equal**? If they indeed are then the chance to win a citation is one in 24,000. What is the chance to win 500 cites out of 350,000? The calculation is slightly more complex than in the militaristic case, but the answer is (see Appendix A) one in $10^{500}$, or, in other words, it is zero. One is tempted to conclude that those forty-four papers, which achieved the impossible, are great.

Recently we discovered [5] that copying from the lists of references used in other papers is a major component of the citation dynamics in scientific publication. This way a paper that already was cited is likely to be cited again, and after it is cited again it is even more likely to be cited in the future. In other words, "unto every one that hath shall be given, and he shall have abundance"[6], [7]. This phenomenon is known as "Matthew effect" [6], "cumulative advantage" [8], or "preferential attachment" [9].

The effect of citation copying on the probability distribution of citations can be quantitatively understood within the framework of **the model of random-citing scientists (RCS)**, which is as follows. When a scientist is writing a manuscript he picks up $m$ random articles[1], cites them, and also copies some of their references, each with probability $p$.

This model resembles a couple of other models [8], [9], [13], [14] (see Appendix B for the key differences[2]), and can be easily solved using methods developed to deal with multiplicative stochastic processes [8], [14].

---

[1]The analysis presented here also applies to a more general case when $m$ is not a constant, but a random variable. In that case $m$ in all of the equations that follow should be interpreted as the mean value of this variable.

[2] These models, though introduced prior to the RCS, are more complicated and difficult to understand for a non-expert reader. This is why discussion of them is moved into Appendix.

The evolution of the citation distribution (here $N_K$ denotes the number of papers that were cited $K$ times, and $N$ is the total number of papers) is described by the following rate equations:

$$\frac{dN_0}{dN} = 1 - m \times \frac{N_0}{N},$$

$$\frac{dN_K}{dN} = m \times \frac{(1 + p(K-1))N_{K-1} - (1 + pK)N_K}{N}, \quad (1)$$

which have the following stationary solution:

$$N_0 = \frac{N}{m+1}; \; N_K = \frac{1 + p(K-1)}{1 + 1/m + pK} N_{K-1}. \quad (2)$$

For large $K$ it follows from (2) that:

$$N_K \sim 1/K^g;$$
$$g = 1 + \frac{1}{m \times p}. \quad (3)$$

Citation distribution follows a power law, empirically observed in [10], [11], [12].

A good agreement between the RCS model and actual citation data [4] is achieved with input parameters $m = 3$ and $p = 1/4$ (see Figure 1). Now what is the probability for an arbitrary paper to become "renowned", i.e. receive more than five hundred citations? Iteration of Eq. 2 (with $m = 3$ and $p = 1/4$) shows that this probability is one in 600. This means that about 40 out of 24,000 papers should be renowned; ergo, mathematical probability, not genius.

On one occasion [15] Napoleon[3] said to Laplace "They tell me you have written this large book on the system of the universe, and have never even mentioned its Creator." The reply was "I have no need for this hypothesis". It is worthwhile to note that Laplace was not against God. He simply did not need to postulate his existence in order to explain existing astronomical data. Similarly, the present work is not blasphemy. Of course, in some spiritual sense, great scientists do exist. It is just that even if they would not exist, citation data would look the same.

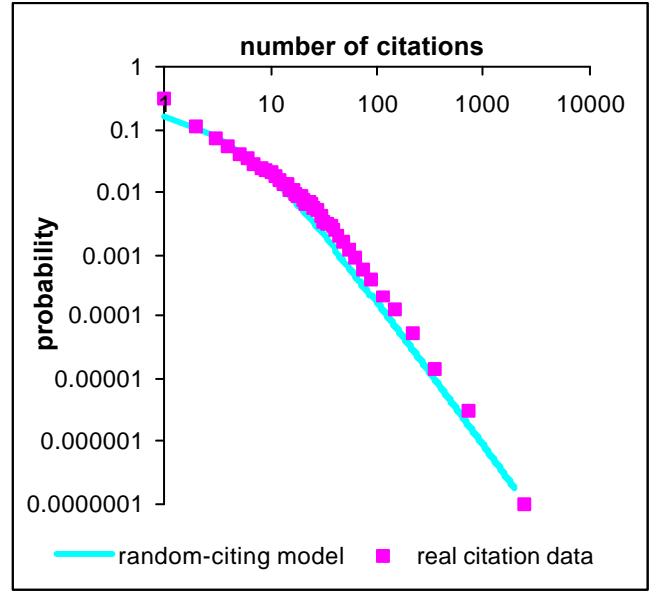

**Figure 1.** Outcome of the model of random-citing scientists (with $m = 3$ and $p = \frac{1}{4}$) compared to actual citation data. Mathematical probability rather than genius can explain why some papers are cited a lot more than the others[4].

## Appendix A

If one assumes that all papers are created equal then the probability to win $m$ out of $n$ possible citations when the total number of cited papers is $N$ is:

$$p = \frac{n!}{m! \times (n-m)!} \times \left(\frac{1}{N}\right)^m \times \left(1 - \frac{1}{N}\right)^{n-m}.$$

Using Stirling formula one can rewrite this as:
$$\ln(p) \approx n \ln(n) - m \ln(m) - (n-m)\ln(n-m)$$
$$- m \ln(N) + (n-m)\ln(1 - 1/N)$$
Substituting $n = 350,000$; $m = 500$; $N = 24,000$ into the above equation we get:
$$\ln(p) \approx -1,282, \text{ or } p \approx 10^{-557}.$$

---

[3] Incidentally, he was the military commander, whose genius was questioned in Ref. [2].

[4] Additional support for the plausibility of this conclusion comes from the findings of Ref. [5] that few citation slips repeat dozens of times, while most appear just once. Certainly no misprint is more seminal than the other.

## Appendix B

In the model introduced by Vazquez [13] a scientist does a recursive bibliography search. Specifically, when he is writing a manuscript, he picks up a paper, cites it, follows its references, and cites a fraction $p$ of them. Afterwards he repeats this procedure with each of the papers that he cited. And so on.

In two limiting cases ( $p = 1$ and $p = 0$ ) the Vazquez model is exactly solvable [13]. Also in these cases it is identical to the RCS model ($m = 1$ case), which in contrast can be solved for any $p$.

Though theoretically interesting, the Vazquez model cannot be a realistic description of the citation process. In fact, the results of Ref. [5] indicate that there is essentially just one "recursion", that is, references are copied from the paper at hand, but hardly followed. To be precise, results of Ref. [5] could support a generalized Vazquez model, in which the references of the paper at hand are copied with probability $p$, and afterwards the copied references are followed with probability $R$ (the "reading" probability introduced in Ref. [5]). However, given the low value of this probability ( $R \approx 0.2$ according to Ref. [5]), it is clear that the effect of secondary recursions on the citation distribution is negligible.

For $p << 1$ effects of second and higher order recursions even in the original Vazquez model are negligible, and it becomes essentially identical to the RCS model. As we find a power law distribution for all non-zero $p$ (see Eq. (3)), this casts doubt on the claim made in [13] that there is a phase transition from power law to exponential distribution around $p \approx 0.4$.

An interesting observation is that in the Vazquez model when $p = 1$ in-component [14] essentially becomes in-degree. This is why Eq.6 of [13] is identical to Eq.59 of [14].

Also Refs [8], [9] by postulating that the probability of paper being cited is somehow proportional to the amount of citations it had already received (no mechanism for this was proposed) were able to explain the power law, which was observed [10], [11], [12] in the distribution of citations.

1. See e.g. W.E. Deming, "Out of the crisis" (MIT, Cambridge, 1986).
2. L. Tolstoy, "War and Peace".
3. E. Garfield, "Citation Indexing" (John Wiley, New York, 1979).
4. SPIRES (http://www.slac.stanford.edu/spires/) data, compiled by H. Galic, and made available by S. Redner:
   http://physics.bu.edu/~redner/projects/citation/
5. M.V. Simkin and V.P. Roychowdhury, "Read before you cite!", cond-mat/0212043.
6. R. K. Merton, "The Matthew Effect in Science", Science, 159, 56 (1968).
7. Gospel according to Mathew, 25:29.
8. D. de S. Price, "A general theory of bibliometric and other cumulative advantage process", Journal of American Society for Information Science, 27, 292 (1976).
9. Barabasi, A.-L. and R. Albert, "Emergence of scaling in random networks", Science, 286, 509 (1999).
10. D. de S. Price, "Networks of Scientific Papers", Science, 149, 510 (1965)
11. Silagadze, Z.K., "Citations and Zipf-Mandelbrot law", Complex Syst. 11, 487 (1997); physics/9901035.
12. Redner, S., "How popular is your paper? An empirical study of citation distribution", cond-mat/9804163; Eur. Phys. J. B 4, 131 (1998).
13. Vazquez, A., "Knowing a network by walking on it: emergence of scaling", cond-mat/0006132; Europhys. Lett. 54, 430 (2001).
14. Krapivsky, P. L. and Redner, S., "Organization of growing random networks", cond-mat/0011094; Phys. Rev. E, 63, 066123 (2001).
15. A. De Morgan, "A budget of paradoxes" (The Open Court Publishing Co., Chicago, 1915). See vol. 2, p.1.